\documentclass{emulateapj}

\usepackage{natbib}
\newcommand{\HII}{{\ion{H}{2}}}

\newcommand{\OII}{[{\ion{O}{2}}]}

\newcommand{\OIIIHb}{[{\ion{O}{3}}]/H$\beta$}
\def\ratioR23{([\ion{O}{2}]~$\lambda$3727 +[\ion{O}{3}]~$\lambda\lambda$4959,5007)/H$\beta$}
\def\R23{${\rm R}_{23}$}

\newcommand{\NII}{[{\ion{N}{2}}]}

\newcommand{\NIIOII}{[\ion{N}{2}]/[\ion{O}{2}]}

\newcommand{\OH}{$\log({\rm O/H})+12$}

\newcommand{\NIIHa}{[\ion{N}{2}]/H$\alpha$}

\def\O4363{[{\ion{O}{3}}]~$\lambda$4363}

\def\L60{L$_{60}$}

\shorttitle{}
\shortauthors{}

\begin{document}

\title{Metallicity Gradients and Gas Flows in Galaxy Pairs}

\author{Lisa J. Kewley,David Rupke, H. Jabran Zahid}
\affil {University of Hawaii}
\email {kewley@ifa.hawaii.edu}

\author{Margaret J Geller}
\affil{Smithsonian Astrophysical Observatory}

\author{Elizabeth J. Barton}
\affil{University of California, Irvine}

\begin{abstract}
We present the first systematic investigation into the metallicity gradients in galaxy close pairs.  We determine the metallicity gradients 
for 8 galaxies in close pairs using \HII\ region metallicities obtained with high signal-to-noise multi-slit observations with  the  Keck LRIS Spectrograph.  
We show that the metallicity gradients in close pairs are significantly shallower than gradients in isolated spiral galaxies such as 
the Milky Way, M83, and M101.  These observations provide the first solid evidence that metallicity gradients in interacting galaxies are
systematically different from metallicity gradients in isolated spiral galaxies.   Our results suggest that there is a strong relationship between metallicity gradients 
and the gas dynamics in galaxy interactions and mergers.   
\end{abstract}


\section{Introduction}
Galaxy interactions and mergers are fundamental to the formation and evolution of galaxies.   
Current N-body/smoothed particle hydrodynamic (SPH) simulations predict that, as a merger progresses, galaxy disks become disrupted by tidal effects, causing large gas inflows into the central regions where kpc-scale starbursts 
may be triggered \citep{Barnes96}, or AGN may be fueled \citep{Springel05c,Hopkins06}.  At very late stages, starburst-or AGN- driven superwinds may drive gas back out of the central regions \citep{Narayanan08}.   

Evidence for major gas inflows in interacting galaxies has been seen in ionized gas \citep{Rampazzo05}, in neutral gas    \citep{Hibbard96,Georgakakis00,Iono05,Emonts06,Cullen07}, and in kinematic studies using absorption lines \citep{Rupke05,Martin06}.   
Whether large-scale gas flows occur may depend on the mass difference between the interacting galaxies \citep[see e.g.,][]{Woods06,Horellou07}. 
In \citet[][]{Kewley06b}, we compared the nuclear gas-phase metallicities of nearby field galaxies with nearby galaxy pairs.  We found that close pairs have systematically lower metallicities than either field galaxies or more widely separated pairs at the same luminosity \citep[see also][]{Lee04}.    Most isolated late-type spiral galaxies, and starburst barred spirals display strong metallicity gradients (see Henry \& Worthey 1999 for a review).   We proposed a merger scenario in which galaxy interactions drive large gas flows towards the central regions, carrying less enriched gas from the outskirts of the galaxy into the central regions, disrupting metallicity gradients and diluting central metallicities. 
 
Subsequent research into central metallicities of interacting galaxies confirms the relationship between galaxy interactions and central metallicities.  \citet{Rupke08} show that merging luminous infrared galaxies have lower central metallicities compared with local isolated galaxies of similar luminosity or mass, suggesting that large interaction-induced gas flows have diluted the central metallicity of luminous infrared galaxies.  Similarly, \citet{Ellison08} and \citet{Michel08} showed that galaxy pairs in the Sloan Digital Sky Survey have lower central metallicities for their stellar mass than isolated galaxies.   Outliers from the standard mass-metallicity relation for star forming galaxies have been shown to be morphologically disturbed \citep{Peeples09,Alonso10}.  Additional investigations suggest a relationship between large-scale environment and metallicity \citep{Cooper08,Ellison09}. 

\citet{Chien07}  investigated the stellar population ages and metallicities of 12 young star clusters in the merging galaxy pair NGC~4676.   They found a relatively flat distribution of oxygen abundances along the northern tail, suggesting efficient gas mixing within the tail.  \citet{Trancho07b}  and \citet{Bastian09} find similarly flat metallicity  distributions in star clusters in the merging galaxies NGC~3256 and the Antennae, respectively.

\begin{deluxetable*}{llllllllll}
\tablecolumns{8}
\tabletypesize{\scriptsize}
\tablecaption{Sample Galaxies\label{Sample_Table}}
\tablehead{ ID & Common & Coordinates & ${\rm M_B}$ & Nuclear  & Hubble & Separation   &  Central    & R$_{25}$     & Gradient \\
                        & Name      &   (J2000)     &                      &  Class\tablenotemark{a}    &  Type\tablenotemark{b}    &   (kpc)\tablenotemark{c}   & log(O/H)$+12$\tablenotemark{a} & (kpc) &   Slope (${\rm R/R}_{25}$)\tablenotemark{d}}
\startdata
1 & UGC 12914 & 00 01 38.3 $+$23 29 01  &  -20.88 & HII  & S(r)cd? pec & 17.4  & 9.08  & 14.69 & -0.104 $\pm 0.048$   \\
2  & UGC 12915 & 00 01 41.9  $+$23 29 45  & -20.09  & ... & S?  & 17.4    & ... & 11.11  &  ... \\
3 & UGC 312 &  00 31 23.9  $+$08 28 01 & -20.53   & HII & SB?  & 21.1   & 8.78 & 11.67  & -0.133$\pm 0.079$  \\
4 & UGC 813 & 01 16 16.4   $+$46 44 25 & -20.44  & HII  & Sb  & 16.5   &  9.00   & 11.16 & -0.233$\pm 0.045$ \\
5 & UGC 816 & 01 16 20.5 $+$46 44 53 & -20.78  & HII &  S?     & 16.5   & 8.91  & 13.33  & -0.252$\pm 0.032$ \\
6 & NGC 3994 &  11 57 36.9 $+$32 16 39 & -19.85  & C & SA(r)c pec & 24.7   & 8.93  & 6.18 &  -0.139$\pm 0.045$  \\
7 & NGC 3995 &  11 57 44.1 $+$32 17 39  &  -20.67 &  HII & SAm pec & 24.7   &  8.75  & 19.40  & -0.339$\pm 0.059$ \\
8 & UGC 12545 &  23 21 41.9 $+$27 04 15  & -20.01  & HII &  SBcd  &  21.9  &  8.88  & 11.81 & -0.381$\pm 0.157$ \\
9   & UGC 12546 & 23 21 41.2 $+$27 05 14  & -20.03  & AMB & Sbc   &  21.9  & 8.91 & 10.64   &  -0.421$\pm 0.094$ \\
\enddata
\tablenotetext{a}{Spectral classes and central metallicities are derived from nuclear long-slit spectral fluxes from \citet{Barton00}.  We use the  \citet{Kewley06} spectral classification scheme and the \citet{Kewley02} [NII]/[OII] metallicity calibration.}
\tablenotetext{b}{Hubble Types are obtained from the NASA Extragalactic Database.}
\tablenotetext{c}{Separation refers to the projected separation of the pair or projected distance to the closest galaxy in the {\it n}-tuple.}
\tablenotetext{d}{Gradient slope is defined in terms of $\frac{\Delta \log({\rm O/H})}{\Delta {\rm R/R}_{25}}$, where ${\rm R}_{25}$ is the B-band isophote at a surface brightness of 25 mag arcsecond$^{-2}$.}
\end{deluxetable*}

Theoretical simulations that include the effect of mergers on the metallicity distribution in galaxies support the picture of major gas inflows induced by the tidal effects of galaxy interactions.  In \citet{Rupke10}, we used N-body/SPH simulations to investigate the theoretical relationship between metallicity gradients and galaxy major mergers.   These simulations are based on close passage, equal-mass disk galaxy merger models with a range of initial geometries and initial pericenters.    We find that between first and second pericenter (i.e. $\sim 1$~Gyr), the metallicity gradient becomes significantly flattened and that the central metallicity becomes diluted by infalling low metallicity gas.

Recently, \citet{Montuori10} used major merger simulations to show that the circumnuclear metallicities become diluted after the first pericenter passage due to major gas inflows.  These model predictions are consistent with observations by \citet{Kewley06b,Rupke08} and with the simulations of \citet{Rupke10}.  Montuori et al. find a second gas dilution peak at final coalescence in major mergers and significant central metallicity dilution in galaxy fly-bys, highlighting the importance of the gas-phase metallicity for tracing the recent merger or interaction history of galaxies.   

 To date, metallicity gradients have been investigated for relatively small numbers of star clusters and \HII\ regions in a small number of interacting galaxies.  In this paper, we present the first systematic investigation of metallicity gradients in interacting galaxies.  We present Keck LRIS spectra for  \HII\ regions in 8 galaxies in close pairs selected on the basis of their position on the luminosity-metallicity relation.   We show that all close pair galaxies have significantly flatter metallicity gradients than metallicity gradients measured in isolated spiral galaxies.  We suggest that flat metallicity gradients provide a `smoking gun' for recent major gas flows in interacting galaxies.  Throughout this paper, we adopt the flat $\Lambda$-dominated cosmology as measured by the WMAP experiment \citep[$h=0.72$, $\Omega_{m}=0.29$;][]{Spergel03}.

\section{Sample Selection and Observations\label{Sample}}

We selected 5 luminous (${\rm M_B}<-20$) sets of close pairs (separation between 15 - 25 kpc) from the \citet{Barton00} galaxy pairs sample.  Our pairs were selected to (a) span a range of positions on the luminosity-metallicity relation in \citet{Kewley06b}, (b) contain at least one galaxy with suitable orientation for metallicity gradient analysis, and (c) to span a range of galaxy properties that may influence metallicity gradients (pair separation, presence of bars, and Hubble type).  In Table~1, we list our sample galaxies and their properties.

We obtained spectra for 12 - 40 star-forming regions in 8 of our close pair galaxies with the Keck Low-Resolution Imaging Spectrograph (LRIS).  Slitmasks were designed using 1" wide slits based on H$\alpha$ images of the target galaxies observed with the Vatican Advanced Technology Telescope (Barton, E. \& Jansen, R., private communication).  We used the 560nm dichroic with the 900 l/mm red grating and the 600 l/mm grism, giving a spectral resolution of 3.5\AA\ FWHM between 3500 - 5500\AA\ and 2.8\AA\ between 5500 - 7000\AA.  

The spectra were reduced using a custom pipeline developed in IRAF and IDL  that includes standard bias subtraction, flatfielding, rectification, and wavelength calibration.  Spectra were flux calibrated using standard stars observed at a similar airmass and time to the slitmask observations.  We derived emission-line fluxes from each spectrum using UHSPECFIT, a spectral line-fitting routine developed by us.  For continuum fitting, UHSPECFIT utilizes code from \citet{Moustakas06} and the \citet{Gonzalez05} evolutionary stellar population synthesis models.    

Emission-line fluxes were corrected for extinction using the \citet{Cardelli89} extinction curve. The majority of our \HII\ region spectra do not have the electron temperature-sensitive auroral [\ion{O}{3}]~$\lambda 4363$ line, which is only produced in high electron temperature, low metallicity regions.  We derived metallicities [in units of \OH] using the strong emission lines.   For the \NII~$\lambda 6584$/\OII~$\lambda 3727$ ratios covered by our sample [log(\NIIOII)$> -1.3$; \OH$>8.4$], the \NIIOII\ ratio is a strong function of metallicity with very little ($<<0.1$~dex) dependence on the ionization state of the gas.  We use the \citet{Kewley02} \NIIOII\ calibration to derive metallicities for the \HII\ regions in our sample, after removal of spectra potentially contaminated by an AGN or other non-thermal sources according to the \citet{Kewley06} classification scheme.   More details on the spectroscopic pipeline, UHSPECFIT, emission-line flux measurements and classification are provided in \citet{Rupke10b}.

\begin{figure*}[!t]
\epsscale{0.9}
\plotone{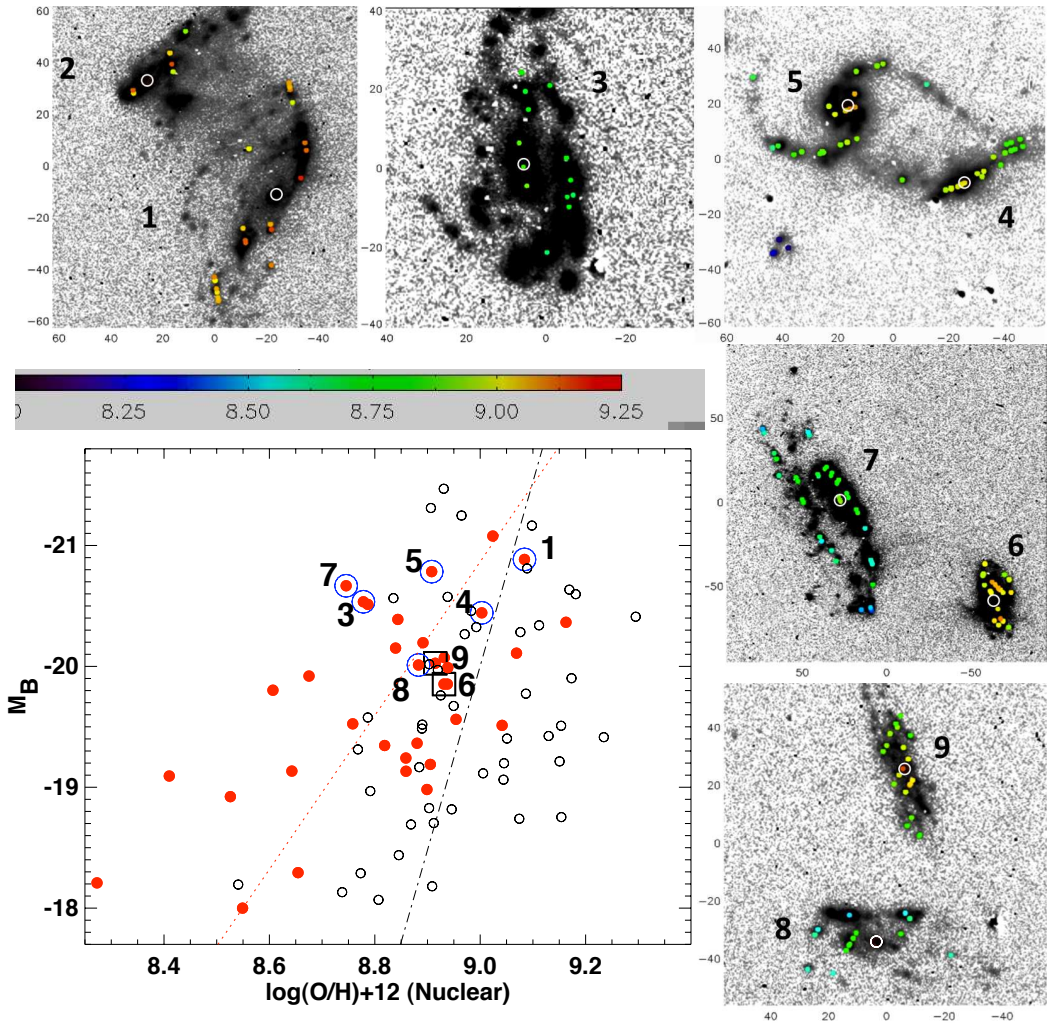}
\caption[f1.eps]{The B-band luminosity-metallicity (L-Z) relation from \citet{Kewley06b} for the close pair members (red) compared with isolated field galaxies from the Nearby Field Galaxy Survey (unfilled circles).  All metallicities in the L-Z relation are calculated from spectroscopy with a central long-slit aperture that covers approximately 10\% of the galaxy B-band light ($\sim 1~$kpc).  The 8 close pairs with metallicity gradients (Table~1) are indicated on the L-Z relation as star forming galaxies (circle outlines) and composite or ambiguous galaxies (square outlines). Top and right panels indicate the location and metallicity of the star-forming regions in each galaxy on H$\alpha$ images.  Metallicity colors are shown in the legend bar in units of \OH\ and the coordinates for each panel are given in units of arcseconds.  Unfilled white circles indicate the galaxy center determined using 2MASS images.  \label{LZ_fig1}}
\end{figure*}

 \citet{Kewley08} show that large discrepancies exist among metallicities derived using different strong-line calibrations.  These discrepancies are systematic, therefore relative metallicities such as gradients may be reliably compared if the same calibration is used.  In addition, we have verified that we obtain the same results if independent metallicity calibrations based on different line ratios are applied such as the \OIIIHb\ $/$ \NIIHa\ calibration by \citet{Pettini04}, and the \R23\ calibrations by \citet{Kobulnicky04} and \citet{McGaugh91}.

\section{Analysis\label{Analysis}}

In Figure~\ref{LZ_fig1}, we show the position of our sample pairs on the luminosity-metallicity relation (where possible) and the spatial distribution of metallicities for each galaxy (colored circles).   Galaxy ID numbers are given in Table~\ref{Sample_Table}.  Spatial distributions of metallicities range from flat (e.g., NGC~3994, \#6) to relatively steep towards the central regions (e.g., UGC~12546, \#9).

\begin{figure}[!t]
\epsscale{1.1}
\plotone{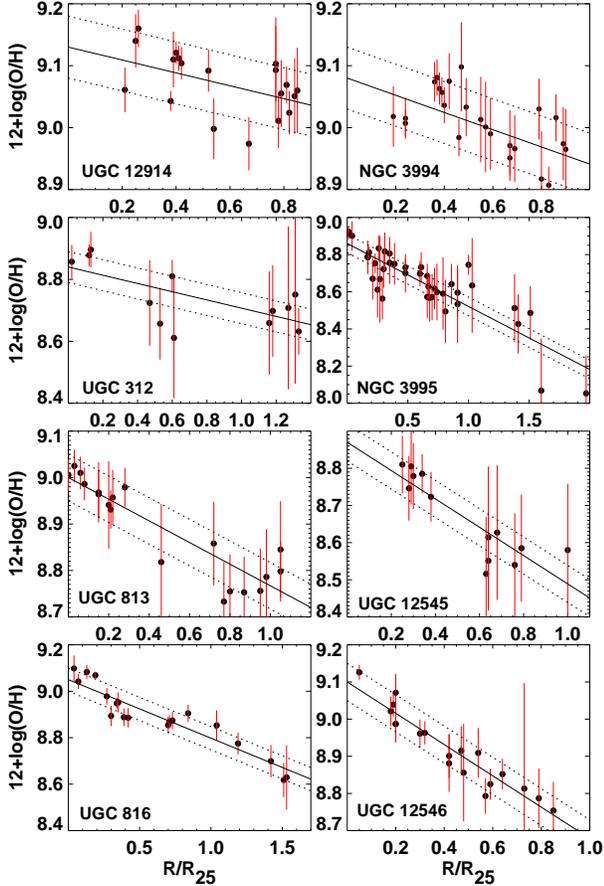}
\caption[f2.eps]{Metallicity gradients derived for our close pair galaxies.  Solid lines indicate the least squares line of best fit to the metallicity gradient and the dotted lines represent the average rms error (0.05~dex) about the line of best fit.  The metallicity gradients are given in terms ${\rm R}/{\rm R}_{25}$ where ${\rm R}$ is the radius and ${\rm R}_{25}$ is the radius of the B-band isophote at a surface brightness of 25 mag arcsecond$^{-2}$.   Similar gradients are obtained using radius in kpc.    The same gradients are obtained within the errors if alternative metallicity calibrations are applied \citep[see][for a detailed comparison]{Rupke10b}.
\label{Grad_example}}
\end{figure}

We measure metallicity gradients for the 8 pair galaxies with at least 10 \HII\ region metallicity measurements.  Galactocentric radii are computed using the direct distance to the galaxy center, where the galaxy center is defined using 2MASS images from a well-defined bulge.  Our distances take inclination into account using the optical diameters, inclinations, and line-of-nodes position angles from HyperLeda \citep{Paturel03}, with double-checks on the position angles by visual inspection of HI maps, where available.   Figure~\ref{Grad_example} shows the metallicity gradients (in ${\rm R}/{\rm R}_{25}$) for the close pair galaxies in our sample.  The \HII\ region radius is given in units of ${\rm R}/{\rm R}_{25}$ where ${\rm R}_{25}$ is the B-band isophote at a surface brightness of 25 mag arcsecond$^{-2}$.

\begin{figure}[!t]
\plotone{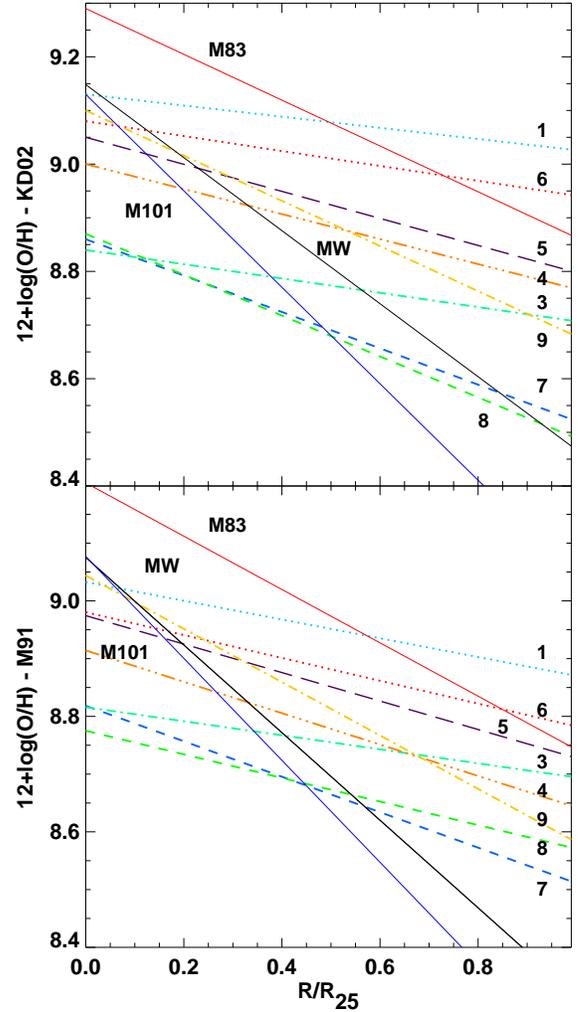}
\caption[f3.eps]{Metallicity gradients for the 8 close pair galaxies in our sample that have metallicities measured for $>10$ HII regions.  Galaxies are labelled according to Table~1.   All metallicity gradients are least-squares fits to HII region metallicities measured using the \citet{Kewley02} [NII]/[OII] calibration (top panel) and the \citet{McGaugh91} \R23\ calibration (bottom panel).   For comparison, we show the metallicity gradients for three isolated spiral galaxies: Milky Way (black), M83 (red), and M101 (blue).    The 8 close pairs have substantially flatter metallicity gradients than the isolated spiral galaxies, providing a smoking gun for major merger-induced gas flows.
\label{gradients}}
\end{figure}

We directly compare the metallicity gradients measured for our close pairs  in Figure~\ref{gradients} for both the  \citet{Kewley02} \NIIOII\ and \citet{McGaugh91} \R23\ calibrations.   For comparison, solid lines indicate the metallicity gradients measured in the isolated spiral galaxies M101, M83 and the Milky Way.   For consistency, we recalculate the \HII\ region metallicities in the isolated spiral galaxies using the \citet{Kewley02} \NIIOII\ and \citet{McGaugh91} calibrations using the \HII\ region emission-line fluxes from \citet{Shaver83} [Milky Way], \citet{Kennicutt03} [M101], and \citet{Bresolin05,Bresolin09} [M83].  The metallicity gradients in these comparison galaxies are representative of isolated spiral galaxies \citep[see review by][]{Henry99}.   We note that we obtain the same results relative to the comparison galaxies if distance between the \HII\ location and the galaxy center is defined in kiloparsecs.

\epsscale{1.1}

The metallicity gradients of all of our close pairs are flatter than the gradients observed in the three isolated spiral comparison galaxies, regardless of calibration used.  The mean gradient in our close pairs is $\frac{\Delta \log({\rm O/H})}{\Delta {\rm R/R}_{25}} =  -0.25 \pm 0.02$ compared with a mean gradient of $\frac{\Delta \log({\rm O/H})}{\Delta {\rm R/R}_{25}} =  -0.67 \pm 0.09$ for the isolated spiral galaxies.  In \citet{Rupke10b}, we show that the mean metallicity gradient for a large sample of 11 isolated spiral galaxies is $\frac{\Delta \log({\rm O/H})}{\Delta {\rm R/R}_{25}}= -0.57 \pm 0.05$.
These results provide the first direct evidence that the metallicity gradients in galaxy pairs are systematically different from gradients in isolated spiral galaxies.  
Recent infall of less enriched gas from the outskirts of the close pair galaxies is the likely culprit. 

 In \citet{Rupke10}, we investigate the effect of gas inflows on metallicity gradients in theoretical merger simulations.  We assume the simplest conditions: no ongoing star formation, equal-mass progenitors and small gas mass fractions.  In these theoretical models, we find a dramatic flattening of the initial radial metallicity gradient between first and second pericenter.  This flattening reflects the effects of gas redistribution over the galaxy disks, including both metal-poor inflow and the growth of tidal tails that carry metals out to large radius.      

 If large merger-triggered gas flows drive the bulk of the inflowing gas into the central kiloparsec, triggering a burst of star formation, a recent nuclear burst of star formation may be associated with a flatter metallicity gradient.   
\citet{Barton03} applied stellar population synthesis models to the B-R colors and Balmer equivalent widths of the galaxy pairs sample to derive the fraction of R-band light originating from a recent burst of star formation, termed the central burst strength.  In \citet{Kewley06b}, we showed that central burst strength correlates strongly with central metallicity in close pairs.  This result is consistent with the theoretical simulations of \citet{Montuori10}, which show that pairs experiencing the strongest bursts of star formation also have the strongest circumnuclear dilution.   If most of the infalling gas is deposited in the central regions, we might expect a correlation between the metallicity gradient slope and central burst strength in our galaxy pairs.  However, we find no statistically significant correlation between the metallicity gradient and the central burst strength (Spearman Rank coefficient of -0.14 with a two-sided probability of obtaining this value by chance of 74\%).   Similarly, we find no statistically significant correlation between the metallicity gradient slope and the presence of blue bulges or pair projected separation.   We note that our gradient sample is limited by the small sample size and that a larger sample is required for more robust conclusions on the relationship between the metallicity gradient slope and galaxy properties.   If the lack of correlation between central burst strength and metallicity gradient slope holds for a larger sample, these results would suggest that (a) the timescale for a central starburst differs from the timescale over which the metallicity gradient remains disrupted, and/or (b) the metallicity gradient slope is strongly influenced by radial gas redistribution along the spiral arms and/or metal outflows, rather than a simple inflow into the central kiloparsec.  

The presence of bars is correlated with efficient gas flows toward the nucleus in late-type spiral galaxies \citep[e.g.,][]{Sakamoto99,Sheth05} and bars may be created in the early stages of a merger \citep{Barnes91}.
Two galaxies in our sample have bars (UGC~312 and UGC~12545).  While UGC~312 has a low central metallicity and relatively flat gradient slope ($\frac{\Delta \log({\rm O/H})}{\Delta {\rm R/R}_{25}} = -0.133$), UGC~12545 has a steeper gradient slope ($\frac{\Delta \log({\rm O/H})}{\Delta {\rm R/R}_{25}} =  -0.381$) that is within the full range of gradient slopes observed in our pairs sample.  It is not clear from these two galaxies whether bars allow more efficient transport of gas towards the central regions during galaxy mergers.
 
We find no correlation between the distance from the nearby field galaxy LZ relation and the close pair metallicity gradient slope.   Galaxy pairs that lie close to the field galaxy LZ relation in Figure~1 may have had high central metallicities [\OH$>9.1$ in the \citet{Kewley02} \NIIOII\ scale] prior to the interaction.  In \citet{Rupke10}, we investigated the theoretical relationship between the metallicity gradient and central metallicities.  We showed that metallicity gradients become disrupted by the interaction, followed by a dilution in the central metallicity.  The observational results presented here support this scenario.

\section{Conclusions \label{Conclusions}}

We present the first systematic study of metallicity gradients in close galaxy pairs.  We find that  {\it all} of our galaxy pairs have flatter metallicity gradients than gradients observed in typical isolated spiral galaxies.  These results show that the low central metallicities seen previously in galaxy pairs and interacting galaxies signal a relatively flat metallicity gradient.  We conclude that large gas inflows are responsible for the disruption and flattening of metallicity gradients in close pair galaxies.    Evidence for gas flows has been difficult to study in the past, requiring integral field or neutral gas velocity maps.   Our results indicate that metallicity gradients provide a unique and feasible method for detecting recent strong tidal gas flows in galaxy interactions and mergers.
Our future work includes comparisons between metallicity gradients and theoretical merger simulations and an extension of our pairs sample to later stage mergers.

\acknowledgments
We thank the referee for comments that improved this paper.  We thank Rolf Jansen for providing H$\alpha$ data used for slitmasks and Li Hsin Chien for preliminary work on some of this data.  The authors thank the indigenous Hawaiian community for use of their sacred mountain.  This research has made use of NASA's Astrophysics Data System Bibliographic Services and the NASA/IPAC Extragalactic Database (NED).


\end{document}